# Intestinal apoptotic changes linked to metabolic status in fasted and refed rats

Caroline Habold[1✉], Charlotte Foltzer-Jourdainne[2], Yvon Le Maho[1],

Jean-Hervé Lignot[1]

[1]CNRS, Centre d'Ecologie et Physiologie Energétiques,

23 rue Becquerel, STRASBOURG cedex 2, F-67087, France ;

[2]INSERM, U381, 3 avenue Molière, Strasbourg, F-67200, France.

✉Correspondence to: Caroline Habold, Centre d'Ecologie et Physiologie Energétiques C.N.R.S. 23 rue Becquerel. 67087 STRASBOURG, France.

E-mail address: caroline.habold@c-strasbourg.fr

Abbreviations: Cdx2: Caudal-related homeodomain protein, RT-PCR: Reverse Transcription – Polymerase Chain Reaction, TGFβ1: Transforming Growth Factor β1, TNFα: Tumor Necrosis Factor α.

**Abstract**

Intestinal apoptosis and expression of apoptosis inducers – the cytokines TNFα, TGFβ1 – and the intestinal transcription factor Cdx2, were studied according to two different metabolic and hormonal phases which characterize long-term fasting: the long period of protein sparing during which energy expenditure is derived from lipid oxidation (phase II), and the later phase characterized by a rise in body protein utilization and plasma corticosterone (phase III). Apoptosis was further studied in 2, 6 and 24h refed rats. Morphological apoptotic events were observed by environmental and conventional scanning electron microscopy and TUNEL test was used to characterize the final stages of apoptotic death. The gene and protein expressions of TNFα, TGFβ1, and Cdx2 were measured. Apoptotic events and TNFα, TGFβ1, and Cdx2 gene and protein expressions did not vary significantly during phase II compared to normally-fed animals. However, phase III fasting induced a delay in intestinal epithelial apoptosis, along with a 92%, 58% and 25% decrease in TNFα, TGFβ1 and Cdx2 mRNAs, respectively. The amounts of TNFα, TGFβ1 and Cdx2 proteins decreased by 70%, 36% and 25%, respectively. Apoptosis was restored rapidly after 2h refeeding following phase III, accompanied by a significant increase in TNFα, TGFβ1 and Cdx2 mRNA and protein levels, compared to the phase III fasting values. The concomitant decreases in cytokines and Cdx2 and in apoptotic cells during phase III suggest the preservation of enterocytes during this critical fasting period in order to optimize nutrient absorption as soon as food is available and thus, to rapidly restore body mass.

Keywords: apoptosis, body reserves depletion, corticosterone, cytokines.

## Introduction

Three distinct metabolic phases have been described during a fast (7, 21, 30, 32). During the first phase (phase I), glycogen stores are completely exhausted and fat stores are progressively used. This phase lasts only a few hours in rats. The second phase (phase II) corresponds to a phase of economy with the mobilization of fat stores for energy expenditure whereas body proteins are efficiently spared. The third phase (phase III) is characterized by an increasing protein catabolism caused by a rise in plasma corticosterone levels. During this later phase, there is a clear change in behavior which promotes food foraging, therefore anticipating a lethal depletion of energy stores (30). In rats, fasting during either phase II or phase III induces a significant decrease in the size of intestinal villi and thus, in the small intestinal mass (13). Also, during the phase of protein sparing (phase II), cell proliferation in crypts and cell migration along the crypt-villus axis decrease, whereas apoptosis at the tip of the villi is maintained (22). In phase III however, when the rat has reached a critical threshold in body reserves utilization, intestinal cell proliferation and migration increase significantly compared to phase II fasted and normally-fed rats. In parallel, we observed a delay in apoptotic events at the tip of the villi (22). The increase in cell proliferation and the preservation of absorptive cells are concomitant with a rise in locomotor activity reflecting the search for food (30) and may prepare the mucosa to nutrient absorption as soon as food is available. The aim of the current work is to further evaluate the intestinal apoptotic changes linked to the metabolic state of fasted rats.

During the normal course of the development of the animal, apoptosis is activated as a response to specific endogenous factors such as TNFα and TGFβ1 (15, 36); the caudal-related homeodomain protein Cdx2 has also been shown to be critical in these processes (3). Apoptosis can be induced by exposure to cytotoxic compounds, hypoxia, viral infection, changing levels of specific hormones, and fasting. A short period of starvation (less than 3 days) has been shown to increase apoptosis in the small intestine mucosa (19, 27, 28). However, other authors (34) have related a decrease in apoptosis following a 3-day starvation period. The effects of longer periods of fasting and of refeeding on apoptotic events in the small intestine remain almost unknown and intestinal apoptosis has never been studied by other authors in relation to whole body metabolism.

In this study, we evaluate the effects of phase II and phase III fasting and of refeeding on apoptotic events in the intestinal epithelium. Apoptosis was studied at the level of cell morphology by Scanning Electron Microscopy (SEM), and cell biology using the TUNEL test. The concomitant variations of proapoptotic cytokines (TNFα, TGFβ1) and of the intestine specific transcription factor Cdx2 were analyzed by RT-PCR in order to determine gene expressions, and by western blot in order to quantify TNFα, TGFβ1 and Cdx2 protein levels. Finally, TGFβ1 and Cdx2 protein distribution along the crypt-villus axis was investigated by immunohistochemical detection.

## Materials and methods

*Animals*

Male Wistar rats weighing 350g were obtained from Iffa-Credo (Lyon, France). The animals were housed individually in leucite cages with a wire mesh floor to minimize coprophagia, and were maintained at 23°C with a 12-hour light period. They had free access to the control diet (A03 pellets from UAR, Epinay-sur-Orge, France) consisting of 23% (by mass) protein, 51.1% carbohydrates, 4.3% fat, 4% cellulose, 5.6% minerals, and 12% water. The rats had free access to water throughout the experiments. They were weighed every day between 9.00 and 10.00 a.m. Our experimental protocol followed the Centre National de la Recherche Scientifique (CNRS) guide for care and use of laboratory animals.

*Experimental procedures*

After a one week acclimatization, normally-fed rats served as control animals (Ctrl, n=15), whereas the other rats were food-deprived. The fasting phases were determined by calculating the specific daily rate of body mass loss dM/Mdt (g/kg/day) for each animal (dM represents the loss of body mass during $dt=t_1-t_0$ and M is the rat body mass at $t_0$). This calculation permitted a daily monitoring of the physiological status for each animal through fasting. With this monitoring, all the animals survived the prolonged starvation procedure and could be successfully refed. The first fasting phase (phase I) lasted only a few hours and was characterized by a rapid decrease in dM/Mdt. The specific daily body mass loss then reached a steady rate (approximately 55g/kg/day) representing phase II and finally, strongly increased which was characteristic of phase III.

The phase II fasting period lasting between one and six days (for 350g rats), a first group of rats was killed in the fourth day in phase II (P2r0, n=15). Three other groups were refed during 2h (P2r2, n=5), 6h (P2r6, n=5), or 24h (P2r24, n=5) following phase II, and then killed. Four additional groups continued fasting until the second day of phase III, reaching on average eight days of fasting. One group was killed in phase III without refeeding (P3r0, n=15), whereas the three others were killed after refeeding for 2h (P3r2, n=5), 6h (P3r6, n=5), or 24h (P3r24, n=5) following phase III.

The animals were killed between 9.00 and 10.00 a.m. The jejunum was removed, weighed, and cut into segments. These segments were then treated separately depending on the analysis considered (CSEM, ESEM, immunohistochemistry, RT-PCR, western blotting).

*Plasma parameters*

Blood samples were collected immediately after sacrifice to measure plasma concentrations of urea and corticosterone in all experimental groups. To assess the metabolic state for each animal, plasma urea was determined with a Urea Nitrogen Kit (Sigma Diagnostics, St Louis, USA), according to the manufacture's guidelines. The plasma concentration of corticosterone was determined with an Enzyme Immunoassay Kit (Assay Designs Inc., Ann Arbor, USA).

*Environmental Scanning Electron Microscopy*

Samples of jejunal mucosa were placed in 1% glutaraldehyde and buffered saline for 30s to block mucus secretion, and then viewed with a Philips XL-30 ESEM. The pressure in the sample chamber was maintained at 5 Torr and the temperature at 4°C, in order to hold a relative humidity rate of 80% at the sample surface.

*Conventional Scanning Electron Microscopy*

Jejunal mucosal samples were fixed for 2h at 4°C in 5% glutaraldehyde in 0.05M cacodylate buffer (pH=7.4), post-fixed 1h in 1% osmium tetroxide, and dehydrated. Pieces were then dried with hexamethyldisilazane. After being mounted on stubs, samples were coated with gold and examined with a Philips XL-30 ESEM.

*TUNEL assay procedure*

The TUNEL assay (in situ Cell Death Detection Kit, fluorescein) was performed according to the manufacturer's specifications (Roche, USA). Each experiment set up by TUNEL reaction mixture without terminal transferase served as negative control. Samples pre-treated with DNAse (2mg/mL) served as positive controls. Sections were examined with a fluorescent microscope (Zeiss Axioplan) equipped with the appropriate filter set (450-590nm bandpass excitation filter). TUNEL-positive cells with fully condensed nuclei were counted in a minimum of 30 villus sections per animal and in 5 animals per group and expressed as a percentage of the total epithelial cells.

*RT-PCR analysis*

Table I

Total RNA from jejunal mucosa was extracted by the method of Chomczynski and Sacchi (8). cDNA was synthesized from 2µg total RNA in 25µL reaction buffer (Finnzymes OY, Finland) containing 1µL of avian myeloblastosis virus reverse transcriptase (Finnzymes OY, Finland), oligo(dT)$_{17}$ primer (50pM, Eurogentec, Belgium) and 0.2mM of deoxynucleotide triphosphate (Promega, France), (60min at 42°C). cDNA was then amplified by Polymerase Chain Reaction (PCR) using specific primers (Table I). The PCR reactions were carried out in 10µL Red'Y'StarMix (Eurogentec, Belgium) diluted to a final volume of 20µL, 50pM of each primer and 1µL of the cDNA mixture. cDNAs were amplified for a determined number of cycles (Table I) as follows: hot starting for 10min at 95°C, denaturation for 45s at 94°C, annealing at 50°C for 45s, and elongation at 72°C for 45s. PCR products were resolved on a 3% agarose gel and visualized by ethidium bromide staining. The staining intensity was evaluated using the Molecular Analyst software (Bio-Rad Laboratories, USA). Results were expressed as relative densitometric units, normalized to the values of a phosphoribosomal protein mRNA used as internal control.

*Western blotting and immunohistochemistry*

Western blotting and immunohistochemistry were conducted as described previously (23). The primary antibodies used were polyclonal rabbit anti-rat TGFβ1 [anti-LC-(1-30)] (generously provided by Dr. K. C. Flanders), monoclonal mouse anti-human Cdx2 (BioGenex, San Ramon, Ca, USA) and polyclonal rabbit anti-human TNFα (Genzyme Diagnostics, Cambridge, USA).

Cdx2 antibody could be detected on paraffin sections after incubation with Alexa $_{488}$-conjugated goat anti-mouse IgG (Molecular Probes), whereas TGFβ1 immunolocalization was performed using the standard extravidin-biotin-peroxidase complex technique and diamino-benzidine coloration.

*Statistical analysis*

Data are presented as mean values +/- SEM. Statistical comparisons of experimental data were performed by one-way analysis of variance (ANOVA) and Tukey post-hoc test by using the software Sigmastat (Jandel). The level of statistical significance was set at $P<0.05$.

## Results

*Evolution of the metabolic and hormonal status*

Table II Plasma urea concentration (Table II) did not vary between control, phase II fasted rats and refed rats following phase II. A phase III fast induced a 3.7-fold significant increase in urea concentration. Uremia then decreased in rats after 6h refeeding following phase III.

Plasma corticosterone concentration (Table II) showed a 13-fold increase in P2r0 fasted animals compared to controls, but it was not significant. After refeeding following a phase II fasting period, plasma corticosterone concentration was lower compared to control values. In phase III fasted rats, plasma corticosterone concentration exhibited a 370-fold significant increase compared to controls, and a 29-fold increase compared to P2r0 fasted rats. Refeeding following phase III induced a rapid decrease in plasma corticosterone.

*Apoptotic events*

Figure 1 Environmental Scanning Electron Microscopy makes it possible to examine fresh, lightly fixed biological specimens (neither dehydrated, nor gold coated) without artifacts linked to sample manipulation and treatment. A morphological study of apoptotic events at the tip of the villi was therefore possible for our 9 experimental groups (figure 1, a to i). These observations were made along the jejunum for 5 animals per group. Apoptotic zones were observed at the villi tips in control (figure 1a) and phase II fasted rats (figure 1b), but very few apoptotic events could be seen in phase III fasted rats (figure 1c) and in rats refed 2 and 6h following phase II (figure 1, d and e). In animals refed 24h following phase II (figure 1f), and in those refed following phase III (figure 1, g to i), the intestinal villi tips looked like those in controls. Observations made at higher magnifications with Conventional Scanning Electron Microscopy (figure 1, j to l) showed apoptotic cells being extruded from the epithelium and losing their microvilli in control (figure 1j) and in phase II fasted rats (figure 1k) but not in phase III fasted rats (figure 1l).

Figure 2 There were no differences in the number of TUNEL-positive cells at the tips of the villi between control and phase II fasted rats (figure 2, a, b and j). Only a few apoptotic cells could be seen in animals refed 2h and 6h after a phase II fast (figure 2, d, e and j). Numerous TUNEL-positive cells could be seen again after 24h refeeding following phase II (figure 2f and j). In phase III, almost no apoptotic cells could be detected at the tip of intestinal villi (figure 2c and j). However, after refeeding following phase III, the apoptotic index was the same as in control animals (figure 2, g to j).

*Changes in TNFα and TGFβ1 gene and protein expressions*

TNFα and TGFβ1 were expressed in the small intestine mucosa (figures 3 and 4). However, TNFα was only faintly expressed, as indicated by the high number of PCR cycles needed to detect its mRNA (Table I).

Figure 3 TNFα transcripts and protein level (figure 3A and B) did not vary during a phase II fast compared to control animals. However, TNFα mRNA levels were very low in animals refed 2h and 6h after phase II compared to controls and phase II fasted rats, but no significant difference in the amounts of TNFα protein could be detected. A phase III fast strongly reduced TNFα gene expression and protein level. After 2h refeeding following phase III, the amount of TNFα mRNA exhibited a 7.5-fold increase compared to the control value, and a 94-fold increase compared to the phase III value. It then decreased during 6h and 24h refeeding following phase III, but was still higher than in control rats. TNFα protein level rose to control value after only 2h refeeding following phase III.

Figure 4 TGFβ1 (figure 4A and B) gene expression and protein level did not vary during a phase II fast compared to controls, but a significant increase was observed between animals refed 2h

following phase II and control and phase II fasted animals. TGFβ1 gene expression and protein level were then lowered to control values in 6h and 24h refed rats following phase II. During phase III fasting, TGFβ1 mRNA level and protein amount were decreased by 58% and 36%, respectively. After refeeding following phase III, TGFβ1 gene expression and protein level increased and rapidly reached the control values.

TGFβ1 was localized in the cytoplasm of intestinal epithelial cells, and particularly in the apical part of the enterocytes (figure 4C). It was observed in the upper third of the intestinal villi in control and refed rats, but was weakly expressed in phase III fasted rats.

*Modulations of Cdx2 expression and localization*

Figure 5

Cdx2 gene expression and protein level (figure 5A and B) did not change during a phase II fast. However, a significant increase was observed after 2h refeeding following phase II. After 6h refeeding, Cdx2 gene expression and protein level were then lowered to control values. A phase III fast induced a 25% decrease in Cdx2 mRNA level and protein amount. After refeeding following phase III, Cdx2 gene expression and protein level increased and rapidly reached the control values.

Cdx2 was localized in the nuclei of intestinal cells and showed a decreasing gradient along the crypt-villus axis (figure 5C). A decrease in the expression of Cdx2 protein was observed in phase III fasted animals compared to controls. The number of strongly labeled nuclei increased in animals refed 2h following either phase II, mainly in the crypt compartment, or phase III, all over the crypt to villus axis.

## Discussion

This study provides evidence that the different apoptotic events described in the small intestinal epithelium during fasting and refeeding appear concomitantly with the different metabolic states of prolonged fasting. During phase II fasting, *i.e.* when the energy requirements are mostly derived from lipid oxidation, changes in intestinal villus cell apoptosis are not obvious compared to normally-fed animals. However, a phase III fast characterized by an increase in protein utilization as a substitute fuel for lipids, induces a down-regulation of apoptotic events accompanied by significant modifications in the expression of the proapoptotic cytokines TNF$\alpha$ and TGF$\beta$1 and of the intestinal specific transcription factor Cdx2. Whereas refeeding after phase II induces a reduction in apoptosis, refeeding following phase III rapidly restores apoptosis at the villi tips with TNF$\alpha$, TGF$\beta$1 and Cdx2 expression, also rapidly normalized. In a previous work (22), we also demonstrated that crypt cell proliferation and migration were decreased during a phase II fast and extensively increased during phase III. All together, these changes could correspond to the preparation of a physiologically functional epithelium able to digest and absorb nutrients at refeeding.

We clearly demonstrate a difference in the occurrence of apoptotic events in the intestinal mucosa according to the different metabolic phases through fasting. During the phase of lipid oxidation (phase II), apoptosis was observed at the villus tips with the same frequency as in normally-fed animals. In phase III however, when the depletion in lipid reserves has reached about 80% of the initial stores, no apoptotic cells were observed at the tips of the villi. This may explain the conflicting results obtained in previous studies. Apoptosis appeared to increase after a very short (1-day) starvation period (27). Because of such a small fasting duration, the rats used in this study and weighing 250-300g were most probably in a phase of lipid oxidation (phase II). During a longer starvation period (3 days), data obtained in younger rats (200-250g) with less adiposity showed on the contrary, a down-regulation of DNA fragmentation (34). These animals may not have been in the same hormonal and metabolic state as in the Iwakiri's study (27), and were likely to be in a phase of protein catabolism (phase III).

From these data, it appeared necessary to further evaluate the effects of phase II and phase III starvation periods on early apoptotic inducers and particularly on TNFα, TGFβ1, and Cdx2, normally present in the gastrointestinal mucosa. TNFα is a cytokine involved in enterocyte apoptosis; mice treated with TNFα show marked villus atrophy (37). High doses of TNFα have been shown to inhibit cell growth, whereas TNFα at low doses promotes epithelial cell proliferation (10, 15, 25, 29, 43). This cytokine is secreted in the intestinal mucosa by immune competent cells present in the *lamina propria*, by intraepithelial lymphocytes (16) and also by enterocytes (39). The decrease in TNFα protein during phase III may induce the delay in enterocyte apoptosis at the tip of the villi thus preventing the intestinal mucosa from further degradation during long-term fasting. In addition, the decrease in TNFα may also favor the increase in crypt cell proliferation previously observed in phase III (22). By its effects on both apoptosis and proliferation, TNFα may contribute to the preservation of the mucosal integrity during phase III. The observed increase in TNFα gene and protein levels 2h after refeeding following phase III, may explain the restoration of epithelial apoptosis.

TGFβ1 is a multipotent cytokine playing an important role in regulating intestinal epithelial cell growth and differentiation (35, 38). It is an inhibitor of crypt cell proliferation (1, 31) and has a proapoptotic role by regulating gene transcription of proteins associated with apoptosis such as Bcl-2 family members (24). In our study, TGFβ1 gene and protein expressions did not vary during the phase II fast. This result is in accordance with the absence of modifications in apoptotic events during a phase II fast. However, despite the marked drop in villus apoptosis after 2h refeeding following phase II, a rapid increase in TGFβ1 mRNA level and protein expression was observed. The proapoptotic effect of TGFβ1 may be blocked by the increase in plasma insulin level (6) occurring after refeeding following phase II (23). In our previous study (22), we demonstrated that refeeding after a phase II fast induces a rapid restoration of villus morphology. At this stage, TGFβ1 could stimulate mucosal repair by increasing extracellular matrix protein synthesis (26), and epithelial cell restitution (9, 11). A similar increase in the gene expression of TGFβ1 has also been reported in aberrant crypt foci and in the surrounding mucosa of refed rats (4). The severe decrease in TGFβ1 gene and protein expressions and the faint epithelial labeling observed in phase III fasting animals could partly explain the disappearance of intestinal apoptotic cells and the increase in crypt cell proliferation (DNA synthesis and mitosis) occurring during this metabolic phase. TGFβ1 mRNA level and protein expression returned to control value after only 2h refeeding following a phase III and was accompanied by the rapid restoration of apoptotic events.

The homeobox gene Cdx2 is known to reduce intestinal cell proliferation, and to stimulate cell differentiation and apoptosis (12, 18, 33, 41, 42). To our knowledge, the effects of short-term and prolonged fasting on Cdx2 *in vivo* are still scanty. Only one study indicates a decrease in Cdx expression in 48h fasted chicks (20). Our results show that Cdx2 is located in the nuclei of crypt cells and of differentiating enterocytes with a decreasing crypt-villus gradient. The amount of Cdx2 protein decreased during the phase III fast and the crypt-villus gradient became barely detectable. This suggests that the absence of Cdx2 in epithelial cells from the villi tips may be involved in the downregulation of apoptosis during phase III and its faint expression in crypt cells may favor the increase in cell proliferation previously observed during this fasting phase (22).

Most of the changes occurring during the phase III fasting period could be linked to the strong increase in plasma corticosterone (this study, 2) concomitant to the decrease in cytokine expression in the jejunum, the increased cell proliferation rate in intestinal crypts and the arrest in apoptotic events at the villi tips. In an early study, the role of glucocorticoids on cytokines secretion was pinpointed as injection of glucocorticoids *in vivo* induced a decrease in TGFβ1 gene expression (40). Glucocorticoids also, negatively regulate TNFα expression in cultured myofibroblats isolated from the jejunum (40) and suppress TNFα secretion by human *lamina propria* lymphocytes (14). Finally, it has been shown that adrenalectomy leads to partial atrophy and disorganization of the villi architecture, associated with a decrease in crypt cell proliferation, and an increase in apoptotic cells at the tip of the villi (17).

Therefore, in light of the data presented here, one can hypothesize that the levels of energy depletion reached by fasting animals are involved in the occurrence of apoptosis in intestinal villi through fasting and after refeeding. The down-regulation of apoptosis during phase III must preserve differentiated epithelial cells and seems strongly correlated with the decreased expression of the homeoprotein Cdx2 and cytokines TNFα and TGFβ1. These changes may lead to the already observed increased cell proliferation rate inducing mucosal repair during the phase III fast, before food becomes available (22). Preservation of absorptive cells and initiation of cell proliferation during phase III fasting is concomitant with a peak of locomotor activity in these animals induced by a rise in plasma corticosterone and reflecting the search for food (5, 30). The unaltered absorption capabilities of the intestine could then permit rapid food assimilation immediately after refeeding and thus, the restoration of the whole body condition. This could be crucial for surviving prolonged fasting, since animals entering phase III, have reached a critical depletion level in their lipid reserves and body proteins.

**Acknowledgments:**

We thank M. Kedinger, C. Domon-Dell and J.N. Freund for helpful discussion, and C. Arbiol and E. Martin for technical help. We are also grateful for K.C. Flanders for providing us the TGFβ1 antibody.

C.H. was recipient of a Nestlé Nutrition grant.

Support of the University Scientific Committee is gratefully acknowledged.

Table I:

Synthetic oligonucleotides and experimental conditions used in reverse transcription-polymerase chain reaction analysis

| *Gene* | *Position* | *Size (bp)* | *Sequence* | *Cycling number* |
|---|---|---|---|---|
| Ribosomal phosphoprotein | 340-360<br>847-828 | 486 | 5'-GTTCACCAAGGAGGACCTCA-3'<br>3'-AGACACCTCTGCCTAATGTG-3' | 23-25 |
| Cdx2 | 578-600<br>649-672 | 250 | 5'-CCCAGCGGCCAGCGGCGAAACCT-3'<br>3'-TATTTGTCTTTTGTCCTGGTTTTC-5' | 28 |
| TGFβ1 | 731-752<br>1415-1392 | 661 | 5'-GAAGTCACCCGCGTGCTAATGG-3'<br>3'-GGATGTAAACCTCGGACCTGTGTG-5' | 28 |
| TNFα | 3033-3055<br>3178-3155 | 122 | 5'-TTCTGTGAAAACGGAGCTAAAC-3'<br>3'-TTTATTACGACTAAACCACTGGT-5' | 37 |

Table II:

Plasma urea and corticosterone concentrations in control, fasted and refed rats

| | *Urea g.L$^{-1}$* | *Corticosterone 10$^{-9}$g.L$^{-1}$* |
|---|---|---|
| Ctrl | 0.189 ± 0.007[a] | 83.98 ± 61.06[a] |
| | | |
| P2r0 | 0.203 ± 0.011[a] | 1077.42 ± 312.80[ab] |
| P2r2 | 0.319 ± 0.025[ab] | 51.90 ± 29.35[a] |
| P2r6 | 0.282 ± 0.014[ab] | 160.34 ± 72.61[a] |
| P2r24 | 0.176 ± 0.015[a] | 607.31 ± 243.19[ab] |
| | | |
| P3r0 | 0.708 ± 0.068[b] | 31101.09 ± 6269.80[b] |
| P3r2 | 0.598 ± 0.067[b] | 708.24 ± 175.74[ab] |
| P3r6 | 0.448 ± 0.009[ab] | 1467.44 ± 523.99[ab] |
| P3r24 | 0.243 ± 0.017[ab] | 678.42 ± 384.57[ab] |

Results are means ± SEM, n=6 rats per group. Within the same column, values with no letter in common are significantly different (P<0.05)

**Figure legends**

Figure 1: ESEM (a to i) and CSEM (j to l) observations of the morphological apoptotic changes in the jejunal mucosa of control (a, j), phase II fasted (b, k), phase III fasted (c, l) rats and in refed rats 2h (d), 6h (e), 24h (f) after a phase II fast and 2h (g), 6h (h), 24h (i) after a phase III fast. V: villus, VT: villus tip, ZA: zone of apoptosis, MO: mucus orifice, VB: villus base, M: mucus, AC: apoptotic cell, MV: microvilli.

Figure 2: Localization of TUNEL-positive cells in the jejunal mucosa of control (a), phase II (b) and phase III fasted (c) rats, and in refed rats 2h (d), 6h (e), 24h (f) after a phase II fast and 2h (g), 6h (h), 24h (i) after a phase III fast. Bar=20µm. LP: *lamina propria*, E: epithelium, VT: villus tip, TP: TUNEL-positive nucleus, NS: non-specific labeling corresponding to red blood cells and immune cells. (j) Percentage of TUNEL-positive cells in intestinal villi. Mean +/- SEM (n=5 animals per group). Values with different letters were significantly different ($P \leq 0.05$).

Figure 3: RT-PCR (A) and western blot (B) analysis of TNF$\alpha$ in Ctrl, P2r0 and P3r0 rats and after refeeding (P2r2, P2r6, P2r24, P3r2, P3r6, P3r24). Mean +/- SEM (n=5 per group). Values with no letters in common were significantly different ($P \leq 0.05$).

Figure 4: RT-PCR (A) and western blot (B) analysis of TGF$\beta$1 in Ctrl, P2r0 and P3r0 rats and after refeeding (P2r2, P2r6, P2r24, P3r2, P3r6, P3r24). Mean +/- SEM (n=5 per group). (C) TGF$\beta$1 immunohistochemical localization in Ctrl, P2r0, P3r0, P2r2 and P3r2 rats. TGF$\beta$1 is localized in the cytoplasm of intestinal epithelial cells and is mainly observed in the intestinal villi of control and refed animals (6h and 24h refeeding data not shown).

Figure 5: RT-PCR (A) and western blot (B) analysis of Cdx2 in Ctrl, P2r0 and P3r0 rats and after refeeding (P2r2, P2r6, P2r24, P3r2, P3r6, P3r24). Mean +/- SEM (n=5 per group). (C) Cdx2 immunohistochemical localization in Ctrl, P2r0, P3r0, P2r2 and P3r2 rats. Cdx2 is located in the nuclei of intestinal cells and is weakly expressed in phase III fasting animals.

Figure 1:

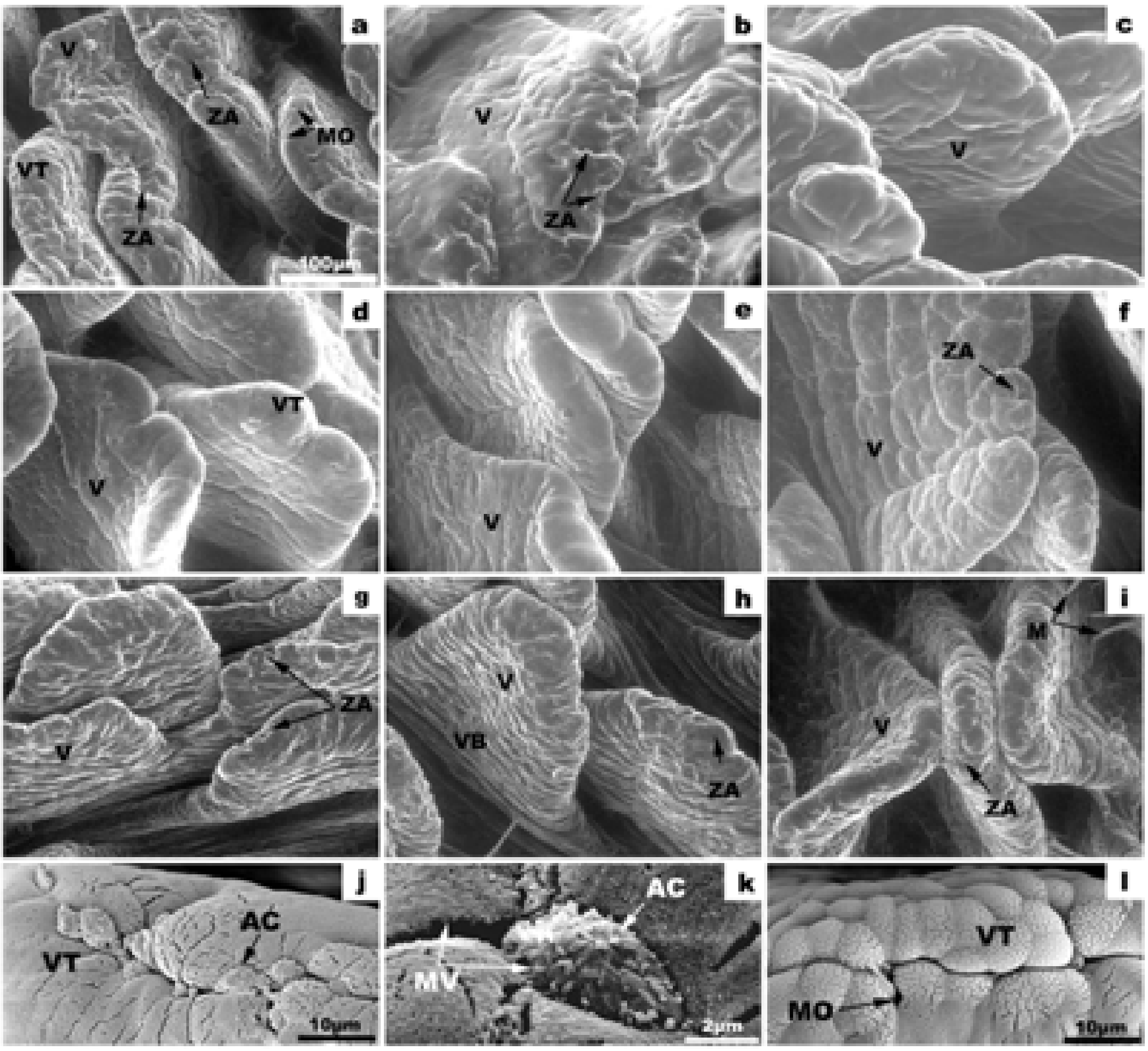

Figure 2:

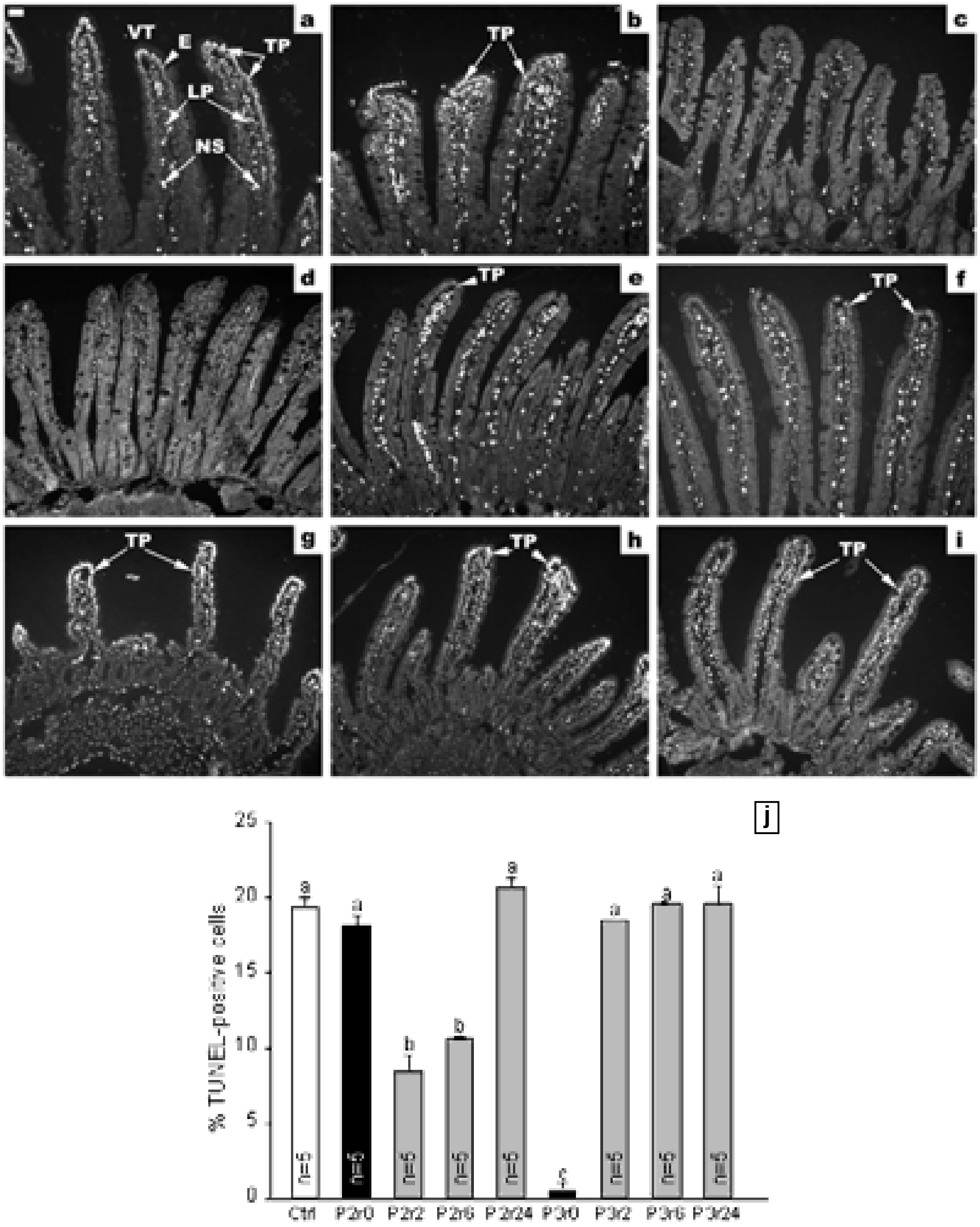

Figure 3:

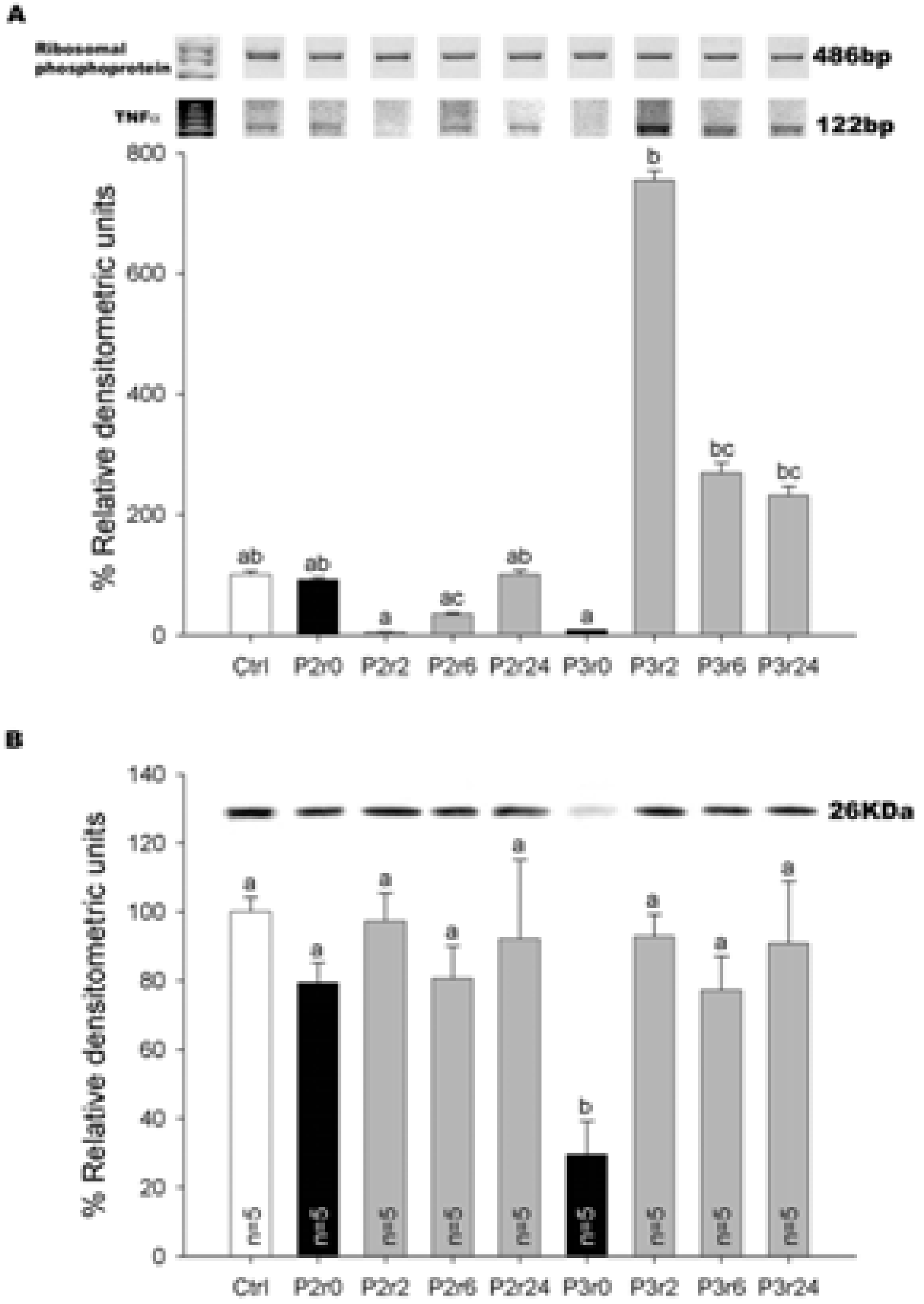
A
Ribosomal phosphoprotein
486bp
TNFα
122bp
% Relative densitometric units
800
600
400
200
0
ab
ab
a
ac
ab
a
b
bc
bc
Ctrl P2r0 P2r2 P2r6 P2r24 P3r0 P3r2 P3r6 P3r24
B
26KDa
% Relative densitometric units
140
120
100
80
60
40
20
0
a
a
a
a
a
b
a
a
a
n=5
Ctrl P2r0 P2r2 P2r6 P2r24 P3r0 P3r2 P3r6 P3r24

Figure 4:

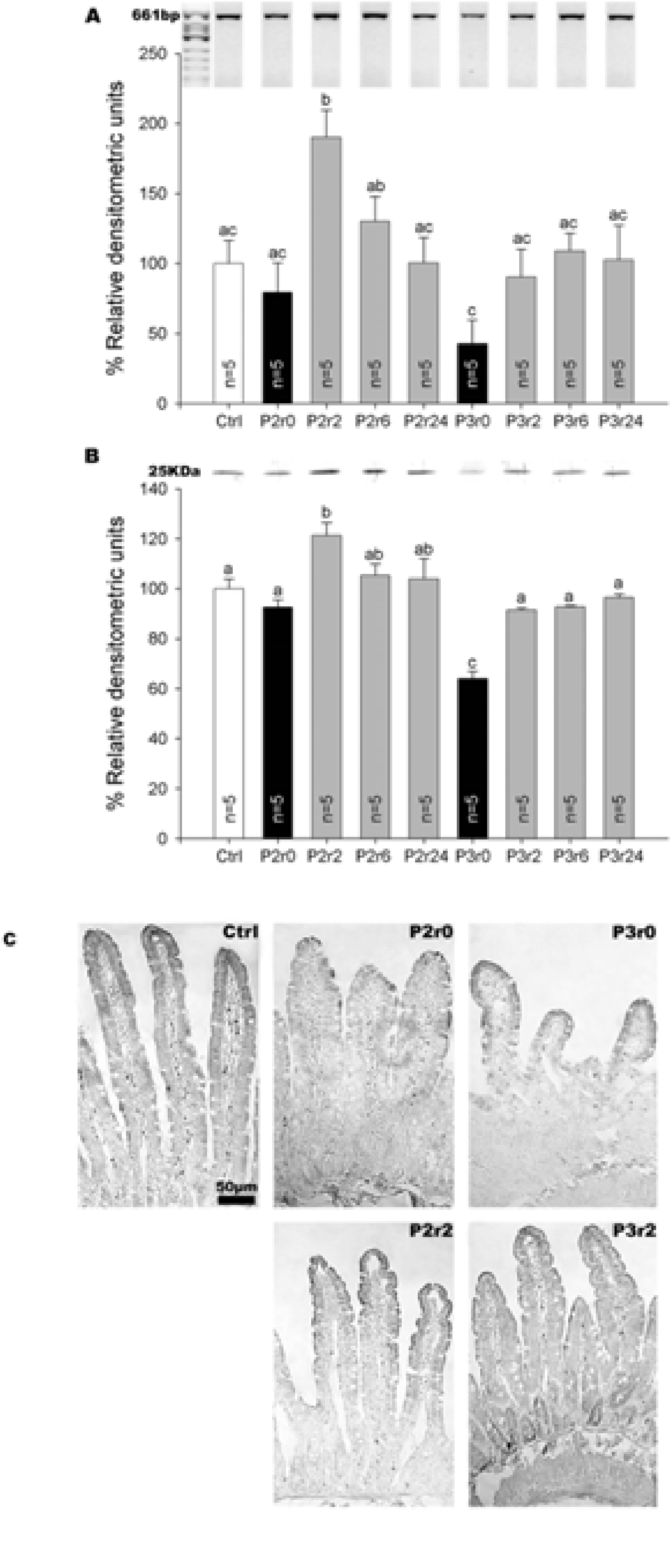

Figure 5:

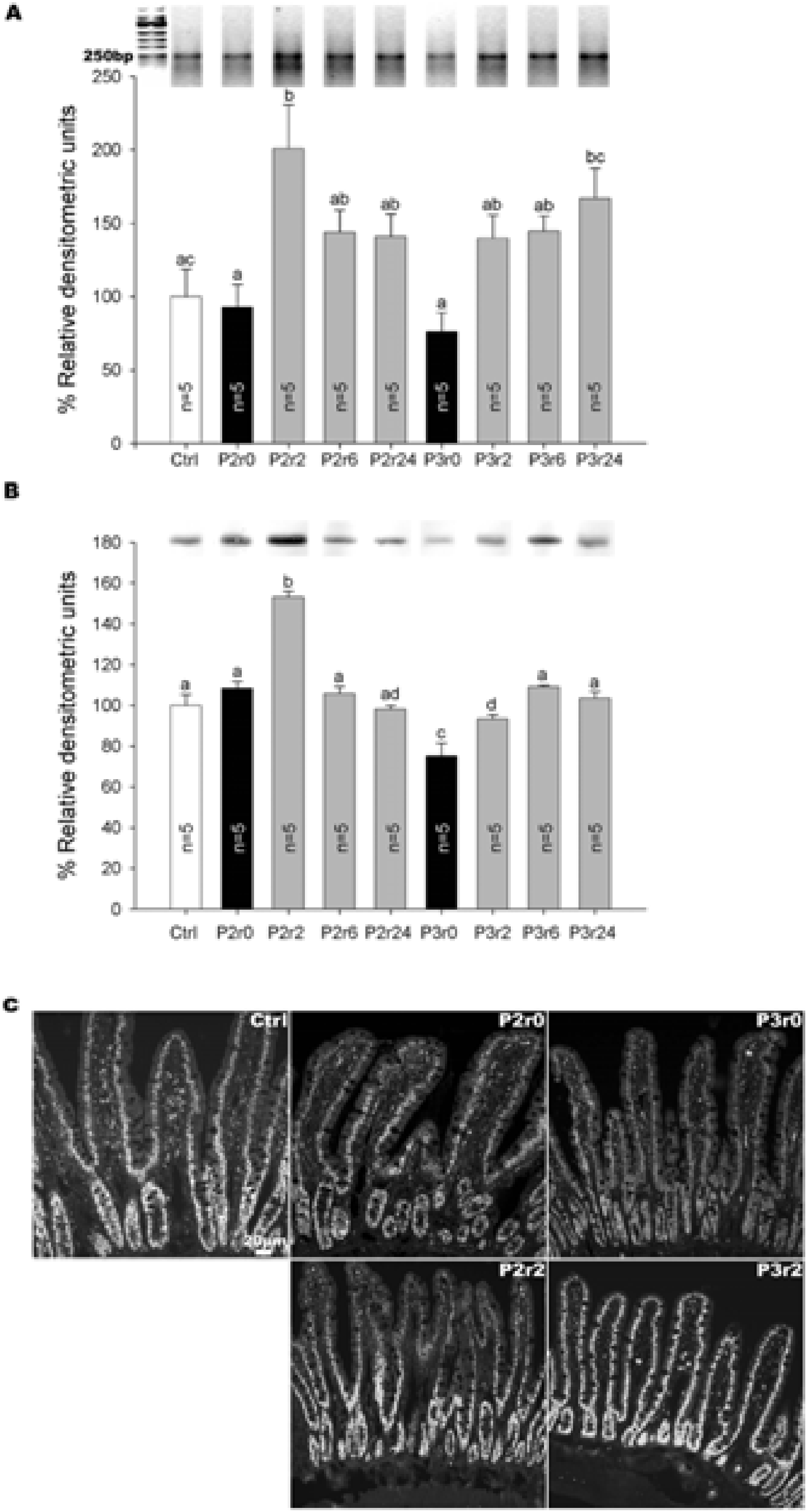